\documentstyle[12pt, a4]{article}
\begin{document}
\author{B. G\"{o}n\"{u}l and M. Ko\c{c}ak
\and Department of Engineering Physics, Faculty of Engineering,
\and University of Gaziantep, 27310 Gaziantep -T\"{u}rkiye}
\title{Explicit solutions for N-dimensional Schr\"{o}dinger
equations with position-dependent mass}
\date{}
\maketitle
\begin{abstract}

With the consideration of spherical symmetry for the potential and
mass function, one-dimensional solutions of non-relativistic
Schr\"{o}dinger equations with spatially varying effective mass
are successfully extended to arbitrary dimensions within the frame
of recently developed elegant non-perturbative technique, where
the BenDaniel-Duke effective Hamiltonian in one-dimension is
assumed like the unperturbed piece, leading to well-known
solutions, whereas the modification term due to possible use of
other effective Hamiltonians in one-dimension and, together with,
the corrections coming from the treatments in higher dimensions
are considered as an additional term like the perturbation.
Application of the model and its generalization for the
completeness are discussed.

\end{abstract}



\section{Introduction}
Gaining confidence from the successful applications \cite{ozer} of
the recently developed simple model \cite{gonul1} in different
fields of physics, we have investigated \cite{gonul2} the relation
between the solutions of physically acceptable effective mass
Hamiltonians proposed in the literature  for the treatment of one
dimensional problems. Using the spirit of the prescription
suggested in \cite{gonul2}, we aim here to tackle the more
difficult problem of generating exact solutions for
position-dependent mass Schr\"{o}dinger equations (PDMSE) in
$N-$dimension, as the most of the related works in the literature
have been devoted to one-dimensional systems except the ones in
\cite{quesne}.

The concept of  PDMSE is known to play an important role in
different branch of physics. This formalism has been extensively
used in nuclei, quantum liquids, $^{3}He$  and metal clusters.
Another area wherein the such concepts provide very useful tool is
the study of electronic properties of many condensed-matter
systems, such as semiconductors and quantum dots. In particular,
recent progress in crystal-growth techniques for producing
non-uniform semiconductor specimens, wherein the carrier effective
mass depends on position, has considerably enhanced the interest
in the theoretical description of semiconductor heterostructures.
It has also recently been signalled in the rapidly growing field
of PT-symmetric or more generally pseudo-Hermitian quantum
mechanics. For an excellent recent review, leading to the related
references, the reader is referred to \cite{quesne}.

In Section 2, the systematic treatment of $N-$ dimensional PDMSE
is presented and closed expressions corresponding to the full wave
function and energy spectrum for exactly solvable potentials are
given. Section 3 contains the application of the model while the
generalization of the formalism is discussed in Section 4.
Concluding  remarks are given in the last section.

\section{Theoretical Consideration}
Tracking down solvable potentials in PDMSE has always aroused
interest. Apart from being useful in understanding  of many
physical phenomena, the importance of searching for them also
stems from the fact that they very often provide a good starting
point for undertaking perturbative calculations of more complex
systems.

As is well known (see, e.g., \cite{gonul2, gonul3}), the general
form of radial PDMSE with Hermitian Hamiltonians in one-dimension
gives rise to
\begin{equation}
-\frac{d}{dz}\left[\frac{1}{M(z)}\frac{d\Phi(z)}{dz}\right]+V^{eff}(z)\Phi(z)=\lambda\Phi(z)~~,
\end{equation}
where the effective potential
\begin{equation}
V^{eff}(z)=V_{0}(z)+U_{\alpha\gamma}(z)=V_{0}(z)-\frac{(\alpha+\gamma)}{2}\frac{M''}{M^{2}}+
(\alpha\gamma+\alpha+\gamma)\frac{M'^{2}}{M^{3}}~~,
\end{equation}
depends on the mass term and ambiguity parameters. Here a prime
denotes derivative with respect to the variable, $M(z)$ is the
dimensionless form of the mass function $m(z)=m_{0}M(z)$ and we
have set $\hbar=2m_{0}=1$. The effective potential is the sum of
the real potential profile $V_{0}(z)$ and the modification
$U_{\alpha\gamma}(z)$ emerged from the location dependence of the
effective mass. A different Hamiltonian leads to a different
modification term. Some of them are the ones proposed by
\cite{bendaniel} BenDaniel-Duke $(\alpha=\gamma=0)$, Bastard
$(\alpha=-1)$, Zhu-Kroemer $(\alpha=\gamma=-1/2)$ and Li-Kuhn
$(\gamma=-1/2, \alpha=0)$.

Considering the works in \cite{gonul2, gonul4}, the radial piece
of PDMSE in arbitrary dimensions for spherically symmetric
potentials and mass functions reads
\begin{equation}
\left\{\frac{d^{2}}{dr^{2}}+\frac{M'}{M}\left(\frac{N-1}{2r}-\frac{d}{dr}\right)-
\frac{L(L+N+2)+(N-1)(N-3)/4}{r^{2}}+M\left[E-V_{eff}(r)\right]\right\}\Psi(r)~~,
\end{equation}
where we assume that $\Psi(r)=F(r)G(r)$ which leads to
\begin{equation}
\frac{1}{M}\left(\frac{F''}{F}+\frac{G''}{G}+2\frac{F'}{F}\frac{G'}{G}\right)-
\frac{M'}{M^{2}}\left(\frac{F'}{F}+\frac{G'}{G}\right)=U_{eff}-E.
\end{equation}
The effective potential in higher dimensions $(N>1)$ now is
transformed to the form
\begin{equation}
U_{eff}(r)=V_{0}(r)+U_{\alpha\gamma}(r)-\frac{M'}{M^{2}}\frac{(N-1)}{2r}+
\frac{L(L+N-2)+(N-1)(N-3)/4}{Mr^{2}}~~,
\end{equation}
in which $L$ is the angular momentum. As the one-dimensional
calculations require $N=1$ and $L=0$, Eq.(5) reduces in this case
to $U_{eff}(r)=V_{eff}(r)=V_{0}+U_{\alpha\gamma}$ as in Ref.
\cite{gonul2}, which provides us a reliable testing ground.

Keeping in mind the spirit of the technique used simply in
\cite{gonul2}, we split Eq. (4) in two parts deviating from the
treatments in \cite{quesne}
\begin{equation}
W^{2}(r)-\left[\frac{W(r)}{\sqrt{M}}\right]'=V_{0}(r)-\varepsilon,
~~W=-\frac{F'}{\sqrt{M}F}~~,
\end{equation}
where $\varepsilon$ is the corresponding energy of the required
quantum state $F_{n}~(n=0,1,2,...)$ for $V_{0}$ which is assumed
in this model as an exactly solvable mass-dependent potential, and
\begin{equation}
\Delta W^{2}(r)-\left[\frac{\Delta
W(r)}{\sqrt{M}}\right]'+2W(r)\Delta W(r)=\Delta V(r)-\Delta E,
~~~~\Delta W(r)=-\frac{G'}{\sqrt{M}G}~~,
\end{equation}
where
\begin{equation}
\Delta V(r)=U\alpha\gamma(r)-\frac{M'}{M^{2}}\frac{(N-1)}{2r}+
\frac{L(L+N-2)+(N-1)(N-3)/4}{Mr^{2}}
\end{equation}
Note that the total energy appearing in (4) is
$E=\varepsilon+\Delta E$ and, in one-dimension the modification
term $\Delta V$ becomes $U_{\alpha\gamma}$ as in Ref.
\cite{gonul2}. This clarifies that the corrections due to the
higher dimensions arise because of the second and third term on
RHS of Eq. (8).

From the present theoretical consideration, Eq. (6) has an
algebraic solution leading to closed analytical expressions for
the wave functions and  energy eigenvalues, hence one needs to
solve Eq. (7) exactly. To proceed further, with the consideration
of relativistic Dirac equations having no ambiguity parameters, we
confidently choose
\begin{equation}
\Delta
W(r)=\frac{(\alpha+\gamma)}{2}\frac{M'}{M^{3/2}}-\frac{(N+2L-1)}{2\sqrt{M}r}~~,
\end{equation}
in which the second term disappears for $N=1$ as in \cite{gonul2}.
Within the frame of Eq. (7), this choice leads us
\begin{equation}
W(r)\Delta
W(r)=\frac{M'}{2rM^{2}}\left[\frac{(\alpha+\gamma)(N-1)}{2}+
(\alpha+\gamma+1)L\right]-\frac{\Delta E}{2}~~,
\end{equation}
that is the main result of the present Letter.

From the definition of the effective potential in Eq. (2), we also
note that the use of Eqs. (7) and (8) naturally restricts  the
choice of some ambiguity parameters yielding different physically
acceptable effective mass Hamiltonians, allowing only
$\alpha=\gamma=0$ (Ben-Daniel Duke Hamiltonian) and
$\alpha=\gamma=-1/2$ (Zhu-Kroemer Hamiltonian) cases. This
observation clarifies that the unperturbed part $(V_{0})$ of the
effective potential in (6) should corresponds to the case
$\alpha=\gamma=0$, having well known solutions in one dimension,
while $\alpha=\gamma=-1/2$ is used to calculate $U_{\alpha\gamma}$
in (8). Obviously, all the corrections coming from the higher
dimensions to the energy and well-behaved wave function terms can
be systematically calculated for a given $M$ with the
consideration of Eqs. (7-10) in the light of corresponding $W$ in
(6).

\section{Application}
Recently, some researches have been devoted to the analysis of the
classification of quantum systems with position-dependent mass
regarding their exact solvability [3-5, and the references
therein]. On a similar basis, Plastino and his co-workers
\cite{plastino} applied an approach within the supersymmetric
quantum mechanical framework, for the case $\alpha=\gamma=0$, to
such systems and succeed to show that some one-dimensional systems
with non-constant mass have a supersymmetric partner with the same
effective mass. They were also able to solve exactly some
particular cases by constructing the superpotential $[W(r)]$ from
the form of the effective mass $[M(r)]$ and generalize the concept
of the shape invariance for these systems.

For illustration, the superpotential expressions given by
\cite{plastino} for the systems having harmonic oscillator and
Morse-like spectra can be easily used in Eq. (6) to serve explicit
expressions for the corrections to the one-dimensional solutions
obtained by considering the Ben-Daniel-Duke effective Hamiltonian
in their \cite{plastino} calculations. This simple investigation
enables us testing our results, because all the corrections should
disappear in case $N=1$ and $\alpha=\gamma=0$ leading to the
expressions in \cite{plastino}. For clarity, this section involves
only the application on the harmonic oscillator system. However,
the generalization of the present model yielding self-consistent
calculations, reproducing $W(r)$ term within the model for any
system of interest, will be discussed in the next section.

According to Ref. \cite{plastino}, $W(r)$ term in Eqs. (6) and
(10) is
\begin{equation}
W(r)=\frac{\omega}{2}\int^{r}\sqrt{M(z)}dz
+\frac{1}{2}\left(\frac{1}{\sqrt{M}}\right)',~~\omega=2\varepsilon_{n=0}~~,
\end{equation}
for the system having harmonic oscillator spectra. Hence, use of
Eqs. (7) through (10) gives
\nonumber
\\
\begin{eqnarray}
\Delta
E=\frac{M'}{rM^{2}}\left[\frac{(\alpha+\gamma)(N-1)}{2}+L(\alpha+\gamma+1)\right]+
\frac{(N+2L-1)\omega}{2r\sqrt{M}}\int^{r}\sqrt{M(z)}dz+
\nonumber
\\
(\frac{1}{\sqrt{M}})'\left\{(\alpha+\gamma)\left[\omega\int^{r}\sqrt{M(z)}dz+
\left(\frac{1}{\sqrt{M}}\right)'\right]+\frac{(N+2L-1)}{2r\sqrt{M}}\right\}~~,
\end{eqnarray}
which is the explicit form of the energy corrections for a given
smooth mass. Clearly, it can be seen that for a constant mass
$M\rightarrow1$, Eq. 12 reduces to $(n+2L-1)\omega/2$  for
arbitrary dimensions \cite{gonul4} while in one dimension it goes
to zero for a non-constant mass in case $\alpha=\gamma=0$
\cite{plastino}. Furthermore, from Eqs. (7) and (9), the
modification term for the corresponding wave function is
\begin{equation}
G(r)=\exp\left(-\int^{r}\sqrt{M(z)}\Delta
W(z)dz\right)=r^{(N+2L-1)/2}M^{-(\alpha+\gamma)/2}.
\end{equation}
As Eq. (6) is analytically solvable having a closed expression for
$W(r)$ given by Eq. (11) reproducing explicit expressions for
$\varepsilon$ and $F$, the corresponding total energy and wave
function can easily be calculated through $E=\varepsilon+\Delta E$
and $\Psi=FG$ for the system of interest with a location dependent
mass. At this stage it is also note that the formalism suggested
here seems superior to the usual treatment in supersymmetric
quantum theory that in principle start with the ground state and
builds up excited state wave functions  by the use of some linear
operators $(A^{\pm})$ whereas there is no such restriction in the
present theory providing flexible investigations.

\section{Discussion}
Although the procedure used in the formalism seems reasonable, the
use of other works as in the previous section for an appropriate
$W(r)$ term to solve Eq. (6) may be seen as a drawback of the
model. To remove this seeming deficiency, we propose here a
unified treatment within the model considering the recent work in
\cite{bagchi}.

Many of the special functions $H(g)$ of mathematics represent
solutions to differential equations of the form
\begin{equation}
\frac{d^{2}H(g)}{dg^{2}}+Q(g)\frac{dH(g)}{dg}+R(g)H(g)=0~~,
\end{equation}
where the functions $Q(g)$ and $R(g)$ are well defined for any
particular function \cite{abramowitz}. Since in this Letter we are
interested in bound state wave functions, we should restrict
ourselves to polynomial solutions of Eq. (14). Bearing in mind Eq.
(14), the substitution of $\Phi(z)=H[g(z)]f(z)$  in Eq. (1) leads
to the second-order differential equation
\begin{equation}
\frac{1}{M}\left(\frac{f''}{f}+\frac{H''g'^{2}}{H}+\frac{g''H'}{H}+2\frac{H'g'f'}{Hf}\right)-
\frac{M'}{M^{2}}\left(\frac{f'}{f}+\frac{H'g'}{H}\right)=V_{eff}-\lambda~~,
\end{equation}
in which primes denote derivatives with respect to $g$ and $z$ for
the functions $H(g)$, $g(z)$ and $f(z)$ respectively. With the
confidence gained by the similarity between Eqs. (15) and (4), one
can safely use the present treatment splitting Eq. (15) in two
pieces
\begin{equation}
W^{2}(z)-\left[\frac{W(z)}{\sqrt{M}}\right]'=V_{0}(z)-\varepsilon,
~~~~W=-\frac{f'}{\sqrt{M}f}~~,
\end{equation}
and
\begin{equation}
\Delta W^{2}(z)-\left[\frac{\Delta
W(z)}{\sqrt{M}}\right]'+2W(z)\Delta W(z)=\Delta V(z)-\Delta E,
~~\Delta W=-\frac{H'g'}{\sqrt{M}H}~~,
\end{equation}
which is similar to Eqs. (6)and (7), where
$\lambda=\varepsilon+\Delta E$ and $V_{eff}=V_{0}+\Delta V$.

After all, it can be clearly seen that Eq. (16) is the one
required for obtaining an explicit expression for $W$ term used in
Eq. (6) corresponding to an exactly solvable system considered in
one-dimension $(\alpha=\gamma=0)$. However, to proceed further,
the functions $f$ and $g$ should be solved as $H$, $Q$ and are $R$
known in principle. Now, equating like terms between the resulting
expression in (15) and (14) gives
\begin{equation}
Q[g(z)]=\frac{1}{g'}\left(\frac{g''}{g'}+\frac{2f'}{f}-\frac{M'}{M}\right),
~~~R[g(z)]=\frac{1}{g'^{2}}\left[\frac{f''}{f}-\frac{M'}{M}\frac{f'}{f}+M(E-V)\right]~~,
\end{equation}
where, from the definition of $Q$,
\begin{equation}
f(z)\approx(\frac{M}{g'})^{1/2}\exp\left[\frac{1}{2}\int^{g(z)}Q(g)dg\right].
\end{equation}
Consideration of Eqs. (15) through (18) suggests a novel
prescription
\begin{equation}
\Delta V(z)-\Delta E=-\frac{g'^{2}}{M}R[g(z)]~~,
\end{equation}
which, for plausible $M$ and $R$ functions, provides a reliable
expression for $g(z)$. It is remarked that in the constant mass
case $M\rightarrow1$ this procedure reduces to the well known
formalism which has been throughly investigated
\cite{bhattacharjie} that, together with \cite{bagchi}, justify
our new proposal in solving PDMSE. The more detailed investigation
of this treatment will be discussed elsewhere.

\section{Concluding Remarks}
In this Letter, a general method has been presented to address the
question of corrections to the solution in one-dimension for a
large class of $N-$ dimensional and exactly solvable PDMSE. We
have also described how to extend the method to the case where the
necessary function $W(r)$ in (6) generating algebraically solvable
potentials in one dimension are present, which initiates
calculations in the model leading to explicit expressions for the
modifications due to both the use of physically plausible
Zhu-Kroemer effective Hamiltonian $(\alpha=\gamma=-1/2)$  and
higher dimensional treatments. The main results are consistent
with the other related works in the literature, which allow a
non-perturbative treatment of these issues.

Although, for clarity we have illustrated  an application of the
method for an easily accessible case of interest, it can be
readily employed in various typical situations. In view of the
importance in calculating such corrections in physics, we believe
that the present model would serve as a useful toolbox to treat
even more realistic situations which now occur in experimental
observations with the advent of the quantum technology.

\end{document}